\documentstyle[aps,prd,epsfig]{revtex}

\newcommand{\be}{\begin{equation}}
\newcommand{\ee}{\end{equation}}
\newcommand{\bea}{\begin{eqnarray}}
\newcommand{\eea}{\end{eqnarray}}
\newcommand{\inli}{\int\limits}

\def\la{\mathrel{\mathpalette\fun <}}
\def\ga{\mathrel{\mathpalette\fun >}}
\def\fun#1#2{\lower3.6pt\vbox{\baselineskip0pt\lineskip.9pt
\ialign{$\mathsurround=0pt#1\hfil##\hfil$\crcr#2\crcr\sim\crcr}}}

\begin{document}

\title{Process $\pi p\to \pi\pi N$ at high energies and moderate
momenta transferred to the nucleon and
the determination of  parameters of the
$f_0(980)$ and $f_0(1300)$}
\author{V.V. Anisovich and A.V. Sarantsev}
\date{\today}
\maketitle

\begin{abstract}
We present the results of simultaneous analysis of the
$S$-wave $\pi\pi$-spectra
in the reactions  $\pi^- p\to (\pi\pi)_S\; n$  at $p_{lab}=38 $ GeV/c
(GAMS) and    $\pi^- p\to (\pi\pi)_S\; n$ at $p_{lab}=18$ GeV/c (E852
Collaboration) at moderate momenta transferred to the nucleon, $|t|\la
1.5$ (GeV/c)$^2$. The $t$-distributions are described by the
reggeized $\pi$- and $a_1$-exchanges provided by
the leading and daughter
trajectories, while the $M_{\pi\pi}$-spectra are determined by a set of
scalar-isoscalar resonances.
With $M_{\pi\pi}$-distributions averaged over $t$-intervals,
we have found several solutions given by different $t$-channel exchange
mechanisms at $|t| \sim (0.5-1.5)$ (GeV/c)$^2$, with resonance
parameters close to each other. We conclude that despite
a poor knowledge of the structure of the $t$-exchange, the
characteristics of resonances such as masses and widths can be reliably
determined using the processes under discussion.
As to pole positions,
 we have found
$(1031\pm 10) - i(35 \pm 6)$ MeV for $f_0(980)$
and $(1315\pm 20) -i(150\pm 30)$ MeV for  $f_0(1300)$.
\end{abstract}

\section{Introduction}

The reactions of meson production in meson--nucleon collisions such as
$\pi p\to \pi\pi N$, $K\bar KN$, $\eta\eta N$ and $KN \to \pi K N$ are
traditionally  a source of information about resonances in the two-meson
spectra, namely, $\pi\pi$, $K\bar K$, $\eta\eta$, $\pi K$, see e.g.
\cite{CERN,Aston,GAMS}. So it would be important to know what resonance
characteristics could be reliably determined from these
reactions and where one may encounter problems.

A set of the $K$-matrix analyses
\cite{km1100,km1500,km1900,AlAn,ufn,YF99,YF} are based on the
three-meson production data in the $p\bar p$ annihilation,
together with data on the two-meson
production in meson-nucleon high energy collisions.
 The GAMS data for the
reaction $\pi^- p\to \pi^0\pi^0 n$  at $p_{lab}=38 $ GeV/c \cite{GAMS},
which represent the production of $\pi^0\pi^0$ at relatively large
momentum transfers, are important for the investigation of 1200--1400
MeV mass region: at $|t|\sim 0.5-1.0$
(GeV/c)$^2$  a distinct peak was seen near 1300 MeV in
$\pi^0\pi^0$ spectra.

The resonance $f_0(1300)$ (denoted as $f_0(1370)$ in the
compilation \cite{PDG}) had been observed in the
analysis of the $\pi\pi$ and $\eta\eta$ spectra obtained from the
annihilation of $p\bar p(at\,rest, liquid\,H_2) \to \pi^0\pi^0\pi^0$,
$\pi^0\pi^0\eta$, $\pi^0\eta\eta$ \cite{VA-PL,VA-PR,Bugg-PR}.  In
the most comprehensive analysis \cite{VA-PR}, where both resonance
production (pole singularities of the amplitude) and  the meson
rescattering in the final state (logarithmic singularities of the
amplitude \cite{log}) were taken into account,
the magnitude for the complex-valued mass was  found to be
$M-i\Gamma/2=(1335\pm 40)-i(127^{+30}_{-20})$ MeV .
In the analyses presented in
\cite{VA-PL,Bugg-PR}, a simplified fitting procedure was carried out,
without accounting for logarithmic singularities: one obtained
$(1340\pm 40)-i(127^{+30}_{-20})$ MeV \cite{Bugg-PR}.
In the Crystal Barrel Collaboration paper  \cite{VA-PL}
the claimed mass was
$1365^{+20}_{-55}$ MeV reflecting an
attempt to make it closer to a scalar resonance
which was then defined at $1430-i125$ MeV
 \cite{Amsler}.  Before that, the existence of a scalar-isoscalar
 resonance was also claimed to be at $1430-i73$ MeV \cite{Kamenski} or
$1420-i110$ MeV \cite{Au}.

But now it became obvious that the mass shift towards higher
value is due to not sufficiently correct account for the interference
"resonance/background": the fact that, just due to a considerable
interference in the $\pi\pi$ $S$-wave the resonance $f_0(1300)$
reveals itself not as a bump or minimum but as a shoulder in the
spectrum, had been specially emphasized in \cite{ufn,VA-PR}.
No visible structure
had been observed in the two-meson spectra from Crystall Barrel
reactions
$p\bar p(at\,rest, liquid\,H_2) \to \pi^0\pi^0\pi^0$,
$\pi^0\pi^0\eta$, $\pi^0\eta\eta$,
although this state was strongly needed for the combined
description of the three-meson Dalitz-plots and the
 two-pion production reactions \cite{CERN,GAMS}:
the combined fits \cite{km1500,km1900,ufn} provide
a strong restriction for the position
of this state --- it could be not higher than 1350 MeV.
The notation $f_0(1370)$ used in the PDG compilation \cite{PDG} is a
tribute to these early (and exaggerated) values \footnote{Note
that in \cite{PDG}, at the  discussion of  the status of $f_0(1370)$,
 the paper \cite{VA-PR} with
the most detailed analysis of this resonance
was not mentioned.}.

There is no common belief in the existence of the $f_0(1300)$,
and its parameters are supposed to be poorly defined: in the
compilation \cite{PDG} the mass and half-width are
quoted as $M\simeq
(1200-1500)$ MeV and $\Gamma/2\simeq (75-125)$ MeV.

The data with a direct evidence for this state
were obtained by the GAMS group  \cite{GAMS} where
the peak associated with $f_0(1300)$ was clearly seen
at large momenta transferred to the target nucleon.
(Note that a hint to a smallness of the background at large
momentum transfers was given by the $K\bar K$ production data
\cite{Polychronakos} where events were collected at $0.2\le|t|\le 0.5$
GeV$^2$ interval, and a strong bump was seen in the mass region $\sim
1300$ MeV).

The GAMS data \cite{GAMS}, where
a strong enhancement in the spectra  was observed in the 1300 MeV mass
region, were included into recent $K$-matrix analyses of the
$IJ^{PC}=00^{++}$-amplitude  \cite{ufn,YF99,YF}.  In \cite{YF}, the
 mass $(1300\pm 20)-i(120\pm 20)$ MeV was found for $f_0(1300)$.
  It became obvious that exchange by
large momenta favours the production of this state, and new
measurement of the $\pi^0\pi^0$ spectra at moderately large $|t|$ in
the reaction $\pi^-p \to \pi^0\pi^0n$ at $p_{lab}=18$ GeV/c performed
by the E852 Collaboration \cite{E852} provided an important
contribution into verification of parameters of this resonance. The
signal from $f_0(1300)$ is clearly seen in the spectra $\pi^0\pi^0$ at
$|t|\sim 0.5-1.5$ (GeV/c)$^2$, for, as was said above, the background
at such momenta transferred is small.

However, the account for the pion-pair-production data in the
$K$-matrix analysis meets with a poor knowledge
of the details of the $t$-channel
exchange mechanism at such momenta. At small momenta,
$|t|\la 0.2$ (GeV/c)$^2$,
the reggeized pion exchange dominates.
At $|t|\sim (0.2-04)$ (GeV/c)$^2$, the behaviour
of the two-pion production cross section with the growth of $|t|$
changes: the decrease of $d\sigma/dtdM_{\pi\pi}$ becomes less steep.
The change of the regime can be due to the onset of different
$t$-channel exchange mechanisms at moderate $|t|$ such as
multi-reggeon rescatterings, say,  $\pi P, \pi PP$, and so on
($P$ denotes the Pomeron) or to the contribution of
the $a_1$ exchange and  related branchings such as $a_1 P, a_1
PP$, etc.

In \cite{km1100}, by performing the $K$-matrix analysis of the GAMS
$\pi\pi$-spectra in the vicinity of the $f_0(980)$, the
$t$-distributions were approximated by the effective pion exchange.
It was supposed that
at small $|t|$ the reggeized pion exchange dominates, while at
increasing $|t|$ the change of regime is accompanied by a change
of sign in the amplitude (recall that the $\pi P$ branching
 changes the amplitude sign). In
the analysis \cite{YF99} of the $\pi\pi$ spectra
in the region of $f_0(980)$ and $f_0(1300)$,
a scenario with a large contribution of the reggeized $a_1$
exchange at $|t|> 0.5$ (GeV/c)$^2$ was realized in the $K$-matrix fit.
The hypothesis that the change of regime in the $t$-distribution at
$M_{\pi\pi} \sim 1000$ MeV is
due to the $a_1$ exchange was also discussed in \cite{Acha}.

To decrease the uncertainties related to a poor
knowledge of the $t$-channel exchange mechanism, the
$M_{\pi\pi}$-distributions averaged over a broad $t$-intervals
$|t_1|\le |t|\le |t_2|$ were used in \cite{km1100} to fit to
data:
\be
\frac{\langle d\sigma \rangle }{dM_{\pi\pi}}=
\inli _{t_1}^{t_2} dt \frac{d\sigma}{dt dM_{\pi\pi}}\ ,
\label{M-dist}
\ee
for the $t$-intervals as follows:
  $0 < |t|<0.2$, $0.3 < |t|<1.0$, $0.35 <
|t|<1.0$, $0.4 < |t|<1.0$,
$0.45 < |t|<1.0$, $0.5 < |t|<1.0$ (GeV/c)$^2$.
The averaged distributions, as one may believe, are not sensitive to
the details of the $t$-distribution, as the averaging over a
broad momentum-transfer interval makes smoother the particularities of
$t$-distributions. Fitting to the spectra
confirmed this statement \cite{YF99,YF}.  In the analysis of
$t$-distributions \cite{YF99},
where, together with pion exchange, the $a_1$
reggeized exchange was included, the parameters of the $f_0(980)$
and $f_0(1300)$ appeared to be  weakly sensitive to different entries
of the $t$-channel exchange mechanism,
thus  giving us a hope that a reliable determination
of resonance pole singularities as well as pole residues
(associated with partial widths) are possible in the framework of
the averaging procedure (\ref{M-dist}).

Recent measurements of
the $M_{\pi\pi}$ spectra in the reaction
$\pi^-p\to \pi^0\pi^0n$ at $|t|<1.5$ (GeV/c)$^2$  \cite{E852}
provide us an opportunity  to enlighten the
$t$-channel mechanism as well as to study to what extent the averaging
of spectra (\ref{M-dist}) makes the extracted resonance parameters
 insensitive to the details of $t$-exchange mechanism.
The present paper is devoted to the consideration of these problems.

As in previous studies \cite{km1100,km1500,km1900,YF99,YF}, we analyse
the $\pi\pi$ spectra in terms of the $K$-matrix amplitude. Because of
that, in Section 2 we recall the necessary
$K$-matix technique formulae. Section 3 presents
 the results of the fit. In Conclusion we summarize our
understanding on the $t$-channel exchange mechanism and recall the
properties of the $f_0(980)$ and $f_0(1300)$ resonances  found in the
$K$-matrix analysis based on the spectra measured by GAMS group
\cite{GAMS} and E852 Collaboration \cite{E852}.

\section{The $K$-matrix amplitude }

In this Section we present the
formulae   for the $K$-matrix analysis of
the $00^{++}$ wave. The given
analysis is a continuation of earlier work
\cite{ufn,YF99,YF}.
In the latter paper \cite{YF} the $00^{++}$ wave
had been reconstructed on the basis of the following data set:\\
(1) GAMS data on the $S$-wave two-meson  production in the reactions
 $\pi p\to \pi^0\pi^0 n$,
$\eta\eta n$ and $\eta\eta' n$
at small nucleon momenta transferred, $|t|<0.2$ (GeV/$c$)$^2$
\cite{GAMS,gams1,gams2};\\
(2) GAMS data on the $\pi\pi$ $S$-wave production in the reaction
 $\pi p\to \pi^0\pi^0 n$
at large momenta transferred,  $0.30<|t|<1.0$ (GeV/$c$)$^2$
\cite{GAMS,gams1};\\
(3) BNL data on $\pi p^-\to K\bar K n$ \cite{bnl};\\
(4)  Crystal Barrel data on
$p\bar p$ (at rest, liquid $H_2$) $\to \pi^0\pi^0\pi^0$,
$\pi^0\pi^0\eta$, $\pi^0\eta\eta$ \cite{VA-PL,cbc}.

Now the experimental basis has been much broadened, and
additional samples of data are included into current analysis of the
$00^{++}$ wave, as follows:\\
(5) Crystal Barrel data on proton-antiproton annihilation in gas:
$p\bar p$ (at rest, gaseous $H_2$ )
$\to \pi^0\pi^0\pi^0$, $\pi^0\pi^0\eta$,
 $\pi^0\eta\eta$ \cite{cbc_new};\\
(6) Crystal Barrel Data on proton-antiproton annihilation
in liquid $H_2$:
$\pi^+\pi^-\pi^0$,
$K^+K^-\pi^0$, $K_SK_S\pi^0$ \cite{cbc_new};\\
(7) Crystal Barrel data on neutron-antiproton annihilation in
liquid deuterium $n\bar p$ (at rest,  liquid
$D_2$) $\to \pi^0\pi^0\pi^-$,
$\pi^-\pi^-\pi^+$,
$K_SK^-\pi^0$, $K_SK_S\pi^-$ \cite{cbc_new};\\
(8) E852 Collaboration data on the $\pi\pi$ $S$-wave production in the
reaction $\pi^-p\to \pi^0\pi^0n$ for nucleon momentum transfers
squared $0<|t|<1.5 \; ({\rm GeV/c}^2$) \cite{E852}.

Below we set out the
$K$-matrix formulae used for the data analysis of the $S$-wave in the
reaction $\pi^-p\to (\pi\pi)_S\, n$.

\subsection{The $K$-matrix scattering amplitude for the
$00^{++}$ partial wave}

The $K$-matrix technique is used for the description
of the two-meson coupled channels:
\be
\hat A= \hat K(\hat I-i\hat{\rho}\hat K)^{-1},
\label{1}
\ee
where $\hat K$ is $n\times n$ matrix ($n$ is the number of channels
under consideration)
and $\hat I$ is the unity matrix.
The phase
space matrix is diagonal:  $\hat{\rho}_{ab}=\delta_{ab}\rho_a$. The
phase
space factor $\rho_a$ is responsible for the threshold singularities
of the amplitude: to keep the amplitude analytical in the
physical region under consideration we use analytical continuation for
$\rho_a$ below threshold.  For example, the $\eta\eta$ phase space
factor $\rho_{\eta\eta}=(1-4m_\eta^2/s)^{1/2}$ is equal to
$i(4m_\eta^2/s-1)^{1/2}$  below  $\eta\eta$ threshold
($s$ is the two-meson invariant energy squared).
To avoid  false singularity in the physical region,
we use for the
$\eta\eta'$ channel the  phase space factor
$\rho_{\eta\eta'}=(1-(m_\eta+m_{\eta'})^2/s)^{1/2}$.

For the multi-meson phase volume in the isoscalar sector, we
use the four-pion phase space defined as either $\rho \rho$
or  $\sigma \sigma$ phase space, where $\sigma$ denotes the $S$-wave
$\pi\pi$ amplitude below  1.2 GeV.  The result does not
depend practically on whether we use $\rho \rho$ or $\sigma \sigma$
state for the description of multi-meson channel: below we provide
formulae and the values of the obtained parameters for the $\rho \rho$
case, for which the fitted expressions  are less cumbersome.

For the $S$-wave amplitude in the isoscalar sector,
we use our standard parametrization  \cite{km1900,ufn,YF}:
\be
K_{ab}^{00}(s)
=\left ( \sum_\alpha \frac{g^{(\alpha)}_a
g^{(\alpha)}_b}
{M^2_\alpha-s}+f_{ab}
\frac{1\;\mbox{GeV}^2+s_0}{s+s_0} \right )\;
\frac{s-s_A}{s+s_{A0}}\;\;,
\label{2}
\ee
with
the following notations for meson states: 1 = $\pi\pi$, 2 = $K\bar K$,
3 = $\eta\eta$, 4 = $\eta\eta'$ and 5 =  multi-meson states
(four-pion state mainly at $\sqrt{s}<1.6\; \mbox{GeV}$).
The $g^{(\alpha)}_a$ is a coupling constant of the bare state
$\alpha$ to the meson channel; the parameters $f_{ab}$ and $s_0$
describe the smooth part of the $K$-matrix elements ($s_0>1.5$ GeV$^2$).
We use  the factor $(s-s_A)/(s+s_{A0})$ to suppress the effect of the
false kinematical singularity at $s=0$ in the amplitude near the
 $\pi\pi$ threshold. Parameters $s_A$ and $s_{A0}$ are kept to be of
the order of $s_A\sim (0.1-0.5)m^2_\pi$ and $s_{A0}\sim (0.1-0.5)$
GeV$^2$ (note that the upper limit of $s_{A0}$ coincides with the
position of the $\rho$-meson left-hand singularity); for these
intervals the results do not depend practically on precise values of
$s_A$ and $s_{A0}$.

For the two-meson states,
$\pi\pi$, $K\bar K$, $\eta\eta$, $\eta\eta'$,
the  phase space matrix elements  are equal to:
\be
\rho_a(s)=\sqrt{\frac{s-(m_{1a}+m_{2a})^2}{s}}\qquad , \qquad a=1,2,3,4
\label{3}
\ee
where $m_{1a}$ and $m_{2a}$ are masses of the pseudoscalars.
The multimeson phase space factor is defined as
\be
\rho_5(s)= \left \{ \begin{array}{cl}
\rho_{51}\;\;\mbox{at} \;\;s<1\;\mbox{GeV}^2,\\
\rho_{52}\;\;\mbox{at} \;\;s>1\;\mbox{GeV}^2,
 \end{array} \right.
\ee
$$
\rho_{51}=\rho_0\int\frac{ds_{1}}{\pi}\int\frac{ds_{2}}{\pi}
 M^2\Gamma(s_{1})\Gamma(s_{2})
\sqrt{(s+s_{1}-s_{2})^2-4ss_{1}}\times
$$
$$
\times s^{-1} [(M^2-s_{1})^2+M^2\Gamma^2(s_{1})
]^{-1}
[(M^2-s_{2})^2+M^2\Gamma^2(s_{2}) ]^{-1} ,
$$
$$
\rho_{52}= \left (\sqrt{\frac{s-16m_\pi^2}{s}}\right )^n\ .
$$
Here $s_1$ and $s_2$ are the two-pion energies
squared, $M$ is $\rho$-meson mass and
$\Gamma(s)$ is its energy-dependent width,
$\Gamma(s)=\gamma\rho_1^3(s)$. The factor $\rho_0$ provides the
continuity of $\rho_5(s)$  at $s=1$ GeV$^2$.
The power parameter $n$ is taken to be 1, 3, 5
for different variants of the fitting;  the results are weakly
dependent on these values (in our
previous analysis \cite{YF} the value $n=5$ was used).

\subsection{The $S$-wave $\pi\pi$, $K\bar K$, $\eta\eta$ and
$\eta\eta'$ production in the high-energy
$\pi p$ collisions}

Here we present formulae for the high-energy $S$-wave
production of $\pi\pi$, $K\bar K$, $\eta\eta$, $\eta\eta'$
at small and moderate momenta transferred to the nucleon.
In \cite{GAMS,E852,gams1,gams2,bnl},
the $\pi p$ collisions were studied at $p_{beam}\sim (15-40)$ GeV/c
(or $s_{\pi N}\simeq 2m_N p_{beam} \sim 30-80 $ GeV$^2$). At such
energies, two pseudoscalar mesons are produced
due to the $t$-channel exchange by  reggeized mesons
belonging to the $\pi $ and $a_1$ trajectories, leading and daughter
ones.

The $\pi$ and $a_{1}$ reggeons have different signatures,
$\xi_\pi =+1$ and $\xi_{a1} =-1$. Accordingly, we
write  the $\pi$ and $a_{1}$ reggeon propagators as:
\be
e^{i\frac{\pi}{2}\alpha_\pi (t)}
\frac{s^{\alpha_\pi (t)}_{\pi N}}{\sin (\frac{\pi}{2}\alpha_\pi (t))}
\quad ,
\qquad
ie^{-i\frac{\pi}{2}\alpha_{a1} (t)}
\frac {s^{\alpha_{a1}  (t)}_{\pi N}}{\cos (\frac{\pi}{2}\alpha_{a1}
(t))}\quad \ .
\label{7}
\ee

Following \cite{syst}, we use for
leading trajectories:
\be
\alpha_{\pi (leading)} (t)\simeq -0.015 +0.72 t , \quad
\alpha_{a1 (leading) } (t) \simeq   -0.10 +0.72 t,
\label{8}
\ee
and for  daughter ones:
\be
\alpha_{\pi ( daughter)} (t)\simeq -1.10 +0.72 t , \quad
\alpha_{a1( daughter)} (t) \simeq   -1.10 +0.72 t\ .
\label{9}
\ee
Here the slope parameters are in GeV.
In the centre-of mass frame, which is the most convenient for the
consideration of  reggeon exchanges, the
incoming particles move along the $z$-axis with momentum $p$.
 In the
leading order of the $1/p$ expansion,
the spin factors for $\pi$ and $a_1$ trajectories read:
\be
\pi-{\rm trajectory}: \qquad
 (\vec \sigma \vec q_\perp),
\label{10}
\ee
$$
a_1-{\rm trajectory}:\qquad
i (\vec \sigma \vec n_z)
$$
where $\vec n_z=\vec p_{beam}/p_{beam}$ and $\vec q_\perp$ is
the momentum transferred to the nucleon ($t\simeq -q^2_\perp$).  The
Pauli matrices $\vec \sigma$ work in the two-component spinor space for
the incoming and outgoing nucleons:  $(\varphi _{out}^* \vec \sigma
\varphi_{in})$ (for  more detail see, for example,
\cite{kaidalov,alkhazov}). Consistent removal from the vertices
(\ref{10}) of the terms decreasing with $p\to \infty$ is necessary for
a correct inclusion of the daughter trajectories which should obey,
similar to the leading ones, the constraints imposed by the $t$-channel
unitarity condition.

In our calculations, we modify conventionally
reggeon propagators in (\ref{7}). We replace:
\be
s_{\pi N}\to \frac{ s_{\pi N}}{ s_{\pi N0}}\ ,
\label{11}
\ee
where the normalization parameter $ s_{\pi N0}$ is of the order
of 4--20 GeV$^2$. To eliminate the poles at $t<0$ we introduce
additional factors into the reggeon
propagators, the Gamma-functions, by replacing in (\ref{7}):
\be
\sin \left (\frac{\pi}{2}\alpha_\pi (t)\right )
\to
\sin \left (\frac{\pi}{2}\alpha_\pi (t)\right ) \;
\Gamma \left (\frac{ \alpha_\pi (t)}{2} +1\right )\ ,
\label{12}
\ee
$$
\cos \left (\frac{\pi}{2}\alpha_{a1} (t)\right )
\to
\cos \left (\frac{\pi}{2}\alpha_{a1} (t)\right )  \;
\Gamma \left (\frac {\alpha_{a1} (t)}{2} +\frac 12\right )\, .
$$
The $K$-matrix amplitude for         the
transitions $\pi R(t)\to \pi\pi$,
$K\bar K$, $\eta\eta$, $\eta\eta'$, $\pi\pi\pi\pi$,
where $R(t)$ refers to reggeon, reads:
\be
\hat A_{\pi R}=\hat K_{\pi R}(\hat I-i\hat{\rho}
\hat K)^{-1},
\label{13}
\ee
where $\hat K_{\pi R}$ is the following vector:
\be
K_{\pi R,b}^{00}
=\left ( \sum_\alpha \frac{G^{(\alpha)}_{\pi R}(t)
g^{(\alpha)}_b}
{M^2_\alpha-s}+F_{\pi R,b}(t)
\frac{1\;\mbox{GeV}^2+s_{R0}}{s+s_{R0}} \right )\;
\frac{s-s_A}{s+s_{A0}}\;\; .
\label{14}
\ee
Here $G^{(\alpha)}_{\pi R}(t)$ and $F_{\pi R,b}(t)$
are the reggeon $t$-dependent form factors.
The following limits are imposed on the form factors:
\be
G_{\pi\pi}^{(\alpha)}(t\to m^2_\pi)=
g^{(\alpha)}_{\pi\pi}\, ,
\qquad F_{\pi
\pi,a}(t\to m^2_\pi)=f_{\pi\pi,a}\ ,
\label{t-limit}
\ee
where
$g^{(\alpha)}_{\pi\pi}$ and $f_{\pi\pi,a}$ enter
the matrix element (\ref{2}).

Different  parametrizations of the form
factor $t$-dependence were investigated in our analysis.
First, the t-dependence of the form factors is
introduced in the exponential form (denoted as
A-parametrization):
\be
G^{(\alpha)}_{\pi R}(t)=g_{\pi R}
\exp \left ( \beta^{(\alpha)}_R (t-m^2_\pi) \right )\ ,
\qquad F_{\pi\pi,a}(t)=f_{\pi\pi,a}
\exp \left ( \gamma_a (t-m^2_\pi)\right )\ .
\label{t-dep_1}
\ee
 Here,
for the sake of simplicity, we have used the same slopes,
$\gamma_\alpha$,  for non-resonance $K$-matrix terms in the channels
$\eta\eta$, $\eta\eta'$ and $\pi\pi\pi\pi$. Also for the
trajectories $a_{1(leading)}$, $\pi_{(daughter)}$, $a_{1(daughter)}$,
the non-resonance couplings were set to be zero.

In the second type of parametrization denoted as B, a more complicated
$t$-dependence has been used for the
$\pi$ trajectory: it is assumed to be a
two-term exponential form for the
form factor:
\bea
G^{(\alpha)}_{\pi\pi}&=&g_{\pi\pi} \left [
\bigg( (1-\Lambda) \exp \left ( \beta_1^{(\alpha)} (t-m^2_\pi) \right )
+ \Lambda \exp \left ( \beta_2^{(\alpha)} (t-m^2_\pi)\right )\right ]
\ .
\label{B1}
\\
G^{(\alpha)}_{\pi\pi}&=&g_{\pi\pi}
\left [ \exp \left ( \beta_1^{(\alpha)} (t-m^2_\pi) \right ) +
\Lambda (t-m^2_\pi) \exp \left ( \beta_2^{(\alpha)} (t-m^2_\pi)
\right ) \right ]\ .
\label{B2}
\eea
The parametrization C assumes a weaker decrease
with $|t|$ for the second term,
that corresponds to the so-called Orear behaviour
\cite{Orear}:
\bea
G^{(\alpha)}_{\pi\pi}&=&g_{\pi\pi}
\left [ (1-\Lambda) \exp \left ( \beta_1^{(\alpha)} (t-m^2_\pi) \right
) + \Lambda \exp
\left (- \beta_2^{(\alpha)} \sqrt{|t-m^2_\pi|}\right ) \right ]\ ,
\label{Orear_1}
\\
G^{(\alpha)}_{\pi\pi}&=&g_{\pi\pi}
\left [ \exp \left ( \beta_1^{(\alpha)} (t-m^2_\pi) \right ) +
\Lambda (t-m^2_\pi) \exp \left (-\beta_2^{(\alpha)}\sqrt{|t-m^2_\pi|}
\right ) \right ]\ .
\label{Orear_2}
\eea
The other form-factor terms are treated in the same way as
in the parametrization A.  As was said above, the change of regime at
$|t|>0.5$ (GeV/c)$^2$ is possible due to multi-Pomeron exchanges, thus
leading to the  Orear behaviour, see \cite{AX} and
references therein.

\section{Results}

In this Section we present the $K$-matrix analysis results
related to the reactions
$\pi^- p\to (\pi\pi)_S\, n$  at $p_{lab}=38 $ GeV/c
\cite{GAMS} and  $\pi^- p\to (\pi\pi)_S\, n$ at $p_{lab}=18$ GeV/c
\cite{E852}.

In the PWA analysis performed by the E852 Collaboration
\cite{E852} two solutions
had been found. We are fitting to the first one which is called in
\cite{E852} a physical solution because of  its
characteristics at the low-mass region.
However, near 1100 MeV both solutions give close
 results,   thus creating a problem of separating these solutions
above 1100 MeV. Therefore, along with fitting to the first
solution, we have performed the analysis where in the mass region
higher than 1100 MeV the points of the second solution are used. It
occured that  fitting to  this modified second solution
 has not led to any qualitative change as compared to the
first solution but a non-significant re-definition
of  parameters for
the $t$-dependence of reggeon form factors. It is  reason for not
presenting  parameters for the modified second solution, and we
restrict ourselves only by the discussion of the results obtained from
fitting to the first E852 solution.

\subsection{The description of the $M_{\pi\pi}$- and
$t$-distributions in the reaction $\pi p\to (\pi\pi)_S\, n$ at
$0<|t|<1.5$ (GeV/$c$)$^2$}

A comparison of the spectra obtained at
 $p_{lab}=38 $ GeV/c
\cite{GAMS} and   $p_{lab}=18$ GeV/c \cite{E852} points to a change
of the $t$-dependence behaviour with energy. This is clearly seen
in Fig.  1, where  the E852 data are plotted in the interval
$|t|=0.3-0.4$ (GeV/c)$^2$ vesus the difference of GAMS spectra for the
intervals $|t|=0.3-1.0$ (GeV/c)$^2$ and $|t|=0.4-1.0$ (GeV/c)$^2$
(unfortunately the E852 data are presented for other $t$ intervals
than those measured by GAMS). A strong difference of spectra is
seen for $M_{\pi\pi}\sim 1100-1350$ MeV, that reveals a significant
contribution of daughter trajectories into formation of
$M_{\pi\pi}$- and $t$-distributions.

The description of data with form factors parametrized in
the form A is shown in Figs.~2 and 3,
and the corresponding $t$-dependence of the $K$-matrix coupling
constants is presented  in Fig.~4 (normalization constant being
$s_{\pi N0}=4$ GeV$^2$).  In this solution,
the $a_1$ exchange is responsible for the peak at 1-GeV region, while
the peak at 1300 MeV at large $|t|$
 is due to the $\pi$ daughter trajectory.
At $|t|$ between 0.1 and 0.4 GeV$^2$, the $a_{1(leading)}$ and
$\pi_{(daughter)}$
 contributions are responsible for a small peak at 1000 MeV
region. For this solution the form factors do not cross the abscissas,
see Fig. 4;
that means the description of spectra is reached in terms of
Regge poles, without Regge branchings.
The description of GAMS data is quite
satisfactory in this approach (see Fig.~3), although certain deviation
is observed at small $|t|$ in the mass region below 1000 MeV. The
$f_0(1300)$ at large $|t|$ is mainly described by the $\pi_{(daughter)}$
trajectory exchange. For this solution the
$a_{1(leading)}$ contribution is rather
large  at small $|t|$ providing noticeable deviation from the one-term
unitarized amplitude.

Futher improvement can be obtained with the form factor
parametrizations
for the $\pi$-trajectory  in the form B: figures
5, 6, 7 demonstrate the results for one of the variants of this
parametrization. For the variant shown in Figs. 5, 6, 7,  which we
denote as B1, we omitted the
$a_{1(daughter)}$ trajectory. The
  $a_{1(leading)}$-exchange is quite large at $|t|\le
0.4$ GeV$^2$. At rather large $|t|$ the  $\pi_{(leading)}$ and
$\pi_{(daughter)}$ trajectories are responsible for the
peak in the 1300 MeV mass region. The $\pi_{(leading)}$-exchange
is also responsible
for the peak at 1000 MeV while $a_{1(leading)}$-exchange becomes here
very small.  For this solution the pion-exchange form factors
for the states
$f_0^{bare}(720)$, $f_0^{bare}(1230)$ and $f_0^{bare}(1600)$
cross the abscissas, thus
corresponding to the $\pi P$ branching effective contribution.
The coupling of
the $f_0^{bare}(1230)$
state grows with $|t|$ due to the increase of  relative weight
of the $f_0(1300)$  at large $|t|$.
However,
the description of the GAMS data  within the parametrization B1
at small $t$-region is not perfect, see
Fig. 6.
Adding the $a_{1(daughter)}$  trajectory leads to a noticeable
improvement of the description.

Adding the $a_{1(daughter)}$  trajectory,
we obtained the solution shown in Figs.~8, 9, 10
(parametrization B2); it
has no visible problems with the description of either E852 or
GAMS data.
For the $\pi_{(leading)}$ exchanges, this solution
is similar to those found in our previous analyses \cite{YF99,YF}
by fitting to GAMS data only:  two resonance couplings cross the
abscissas at moderate $|t|$.

We have also fitted to data under the assumption that
the change of the  $t$-distribution structure at $|t|>0.4$
(GeV/c)$^2$ is due to the onset of the Orear regime, eqs.
(\ref{Orear_1}) and (\ref{Orear_2}).  For this case (parametrization C)
the results are close to those of the B parametrization, so we do not
present here the $M_{\pi\pi}$- and $t$-distributions.

\subsection{Resonance pole positions for the $f_0(980)$ and $f_0(1300)$
states}

Using the found solutions, we have determined
the positions of pole corresponding to the resonances $f_0(980)$
and $f_0(1300)$:
\bea
(1031\pm 10) - i(35 \pm 6 )\;\; {\rm MeV}\ ,
\qquad \qquad
(1315\pm 20) -i(150\pm 30) \;\; {\rm MeV}\ .
\label{sol2}
\eea
The pole for the $f_0(980)$ is under the $\pi\pi$ and  $\pi\pi\pi\pi$
cuts,  the closest physical region  to this pole
is located below the $K\bar K$ threshold (for more details concerning
the determination of sheets, see \cite{YF}).

Recall that in the
previous $K$-matrix analysis \cite{YF} we obtained for
$f_0(1300)$ the mass value $(1300\pm 20)-i(120\pm 20)$ MeV, while for
the $f_0(980)$ it was $(1015\pm 15)-i(43\pm 8)$ MeV. One can see that
the magnitudes quoted in  \cite{YF} and (\ref{sol2}) agree reasonably
with each other.

By fitting to data on the two-meson spectra at
$|t|\sim  0.5-1.0 $ (GeV/c)$^2$, we should  definitely recognize
that our {\it a priory} knowledge about the $t$-channel exchange
mechanism is poor. In the considered $t$-region, together with the
Regge pole terms ($\pi$ and $a_1$ exchanges), the Regge branching
contributions with  additional  pomeron-induced interactions ($\pi
P$, $\pi PP$, or $a_1 P$, $a_1 PP$, etc. $t$-channel exchanges)
are to be significant.
The contribution of Regge branchings is enhanced at moderately
large $|t|$, this circumstance was known long ago, see e.g.
 \cite{AX,AD}.  The presence of a number of terms in the $t$-channel
exchange mechanism at $|t|\ga 0.5$ (GeV/c)$^2$ makes the
model-independent reconstruction of the $t$-channel amplitude
hardly plausible.  Hence a necessity appears to use
at moderately small momentum transfers the
$M_{\pi\pi}$-distributions,  which are not sensitive to the
details of the $t$-channel mechanism.
Let us
stress once again  that, in our opinion, the
$M_{\pi\pi}$-distributions averaged over a broad interval of momentum
transfers do respond to the problem of finding  masses and widths of
the resonances.

\section{Conclusion}

We have performed the fitting to data to determine  parameters of
the $f_0(980)$ and $f_0(1300)$
observed in the $(\pi\pi)_S$ spectra in the
reaction $\pi^-p\to (\pi\pi)_S\, n$
\cite{GAMS,E852} by checking  several hypotheses about
the $|t|$ exchange mechanism.

Concerning the structure of the $|t|$-channel exchange mechanism, one
can see that the E852  data satisfy well the
suggestion about reggeized $\pi$-exchange dominating small
momentum transfers, $|t|<0.2$ (GeV/c)$^2$,
this very mechanism worked at
GAMS energies  as well \cite{km1100}. With the increase of $|t|$, the
change of regime occurs, and the E852 data definitely confirm this.
Yet, the details of the change of regime remain unclear: this may
happen due the inclusion of the $a_1$-exchange, or the branchings $\pi
P$, $a_1 P$ ($P$ is the Pomeron),  or even due to multiple
rescatterings (the Orear regime). The  E852 data reveal
that at
$|t|>0.2 $ (GeV/c)$^2$ the daughter trajectories (pion or $a_1$-meson)
contribute significantly, and the change of the structure
of $|t|$-distributions with energy
 definitely proves it.

The fitting procedure uses the  $M_{\pi\pi}$ spectra which are
averaged over certain intervals of $|t|$.
With different inputs for the $t$-channel exchange mechanism at $|t |
\simeq 0.4$ (GeV/c)$^2$, we have observed a  stability of the
resonance parameters found for $f_0(980)$ and $f_0(1300)$,
and they are close to those obtained in
previous analysis \cite{YF}. So our analysis does not confirm the
statement of the paper \cite{Acha-2} about a strong dependence of
extracted parameters on the details of the $t$-channel exchange
mechanism at $|t|\ge 0.4$ (GeV/c)$^2$.

\section*{Acknowledgement}

We are grateful to A.V. Anisovich, D.V. Bugg, L.G. Dakhno and V.A.
Nikonov for useful and stimulating discussions. The paper is supported
by the RFBR grant No  01-02-17861.

\begin{figure}[hp]
\begin{center}
\epsfig{file=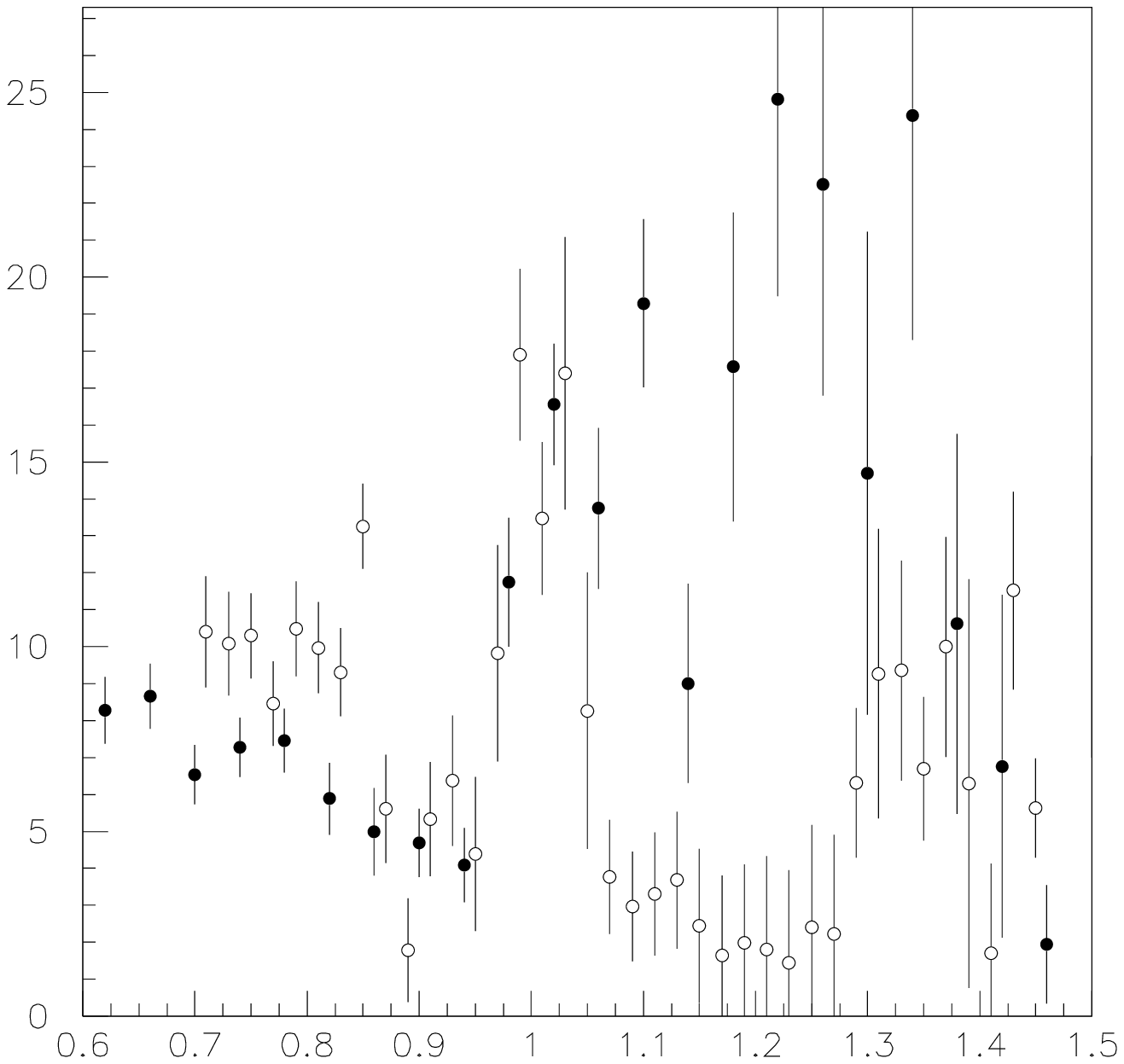,width=14.5cm}\\
Fig. 1. Comparison of the GAMS and E852 data for the
$|t|$-interval $0.3\le |t|\le 0.4$ (GeV/c)$^2$. The full circle are
E852 data and open circles correspond to the subtraction of two
sets of GAMS data, $|t|=0.3- 1.0$ (GeV/c)$^2$ and $|t|=0.4- 1.0$
(GeV/c)$^2$.
\end{center}
\end{figure}

\begin{figure}[hp]
\begin{center}
\epsfig{file=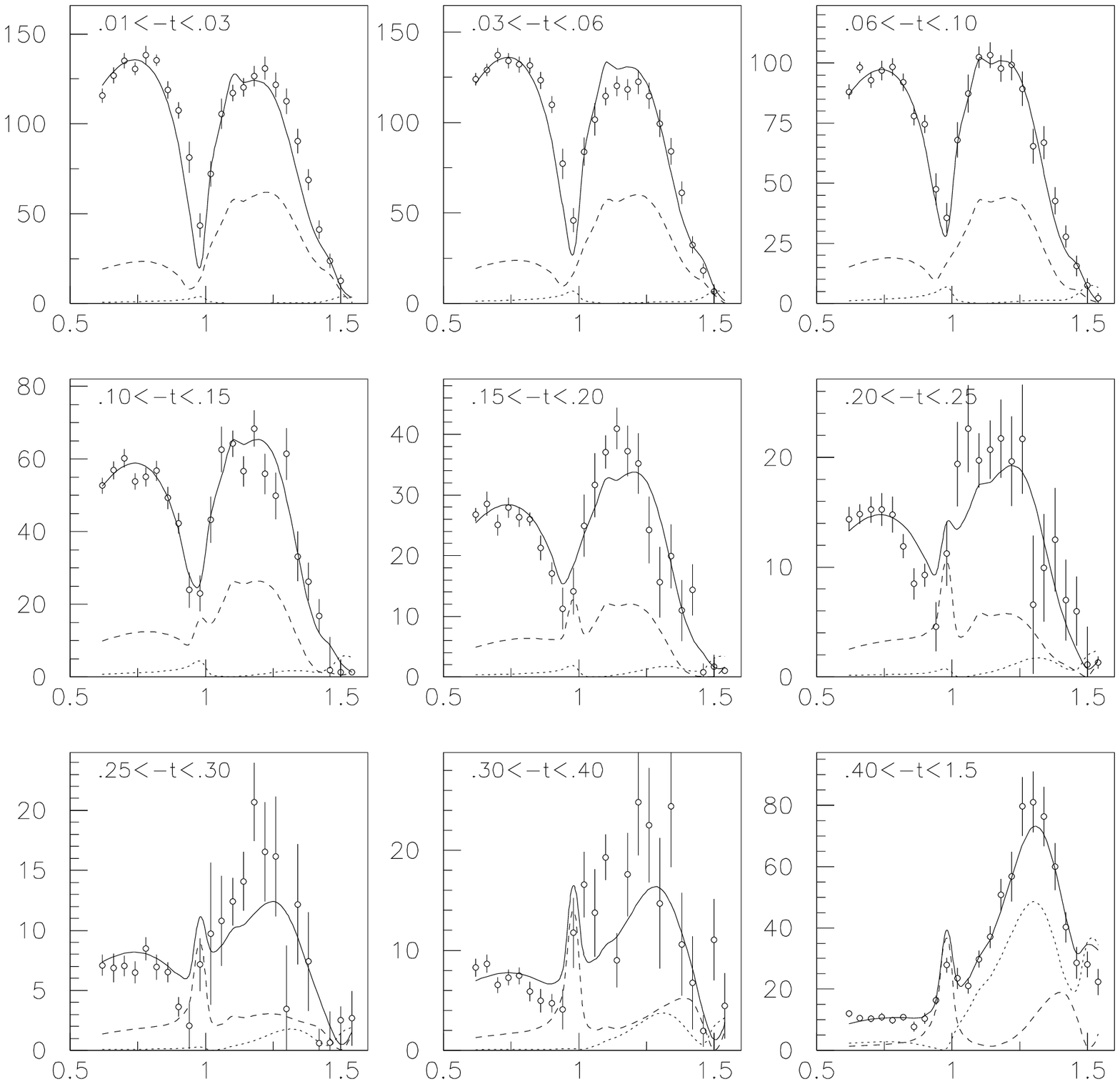,width=14.5cm}\\
Fig. 2. Description of the E852 data in different $t$-intervals for
solution A. Dashed and dotted curves  show the contribution of
 $a_{1(leading)}$ and $\pi_{ (daughter)}$ trajectories, correspondingly.
\end{center}
\end{figure}

\begin{figure}[hp]
\begin{center}
\epsfig{file=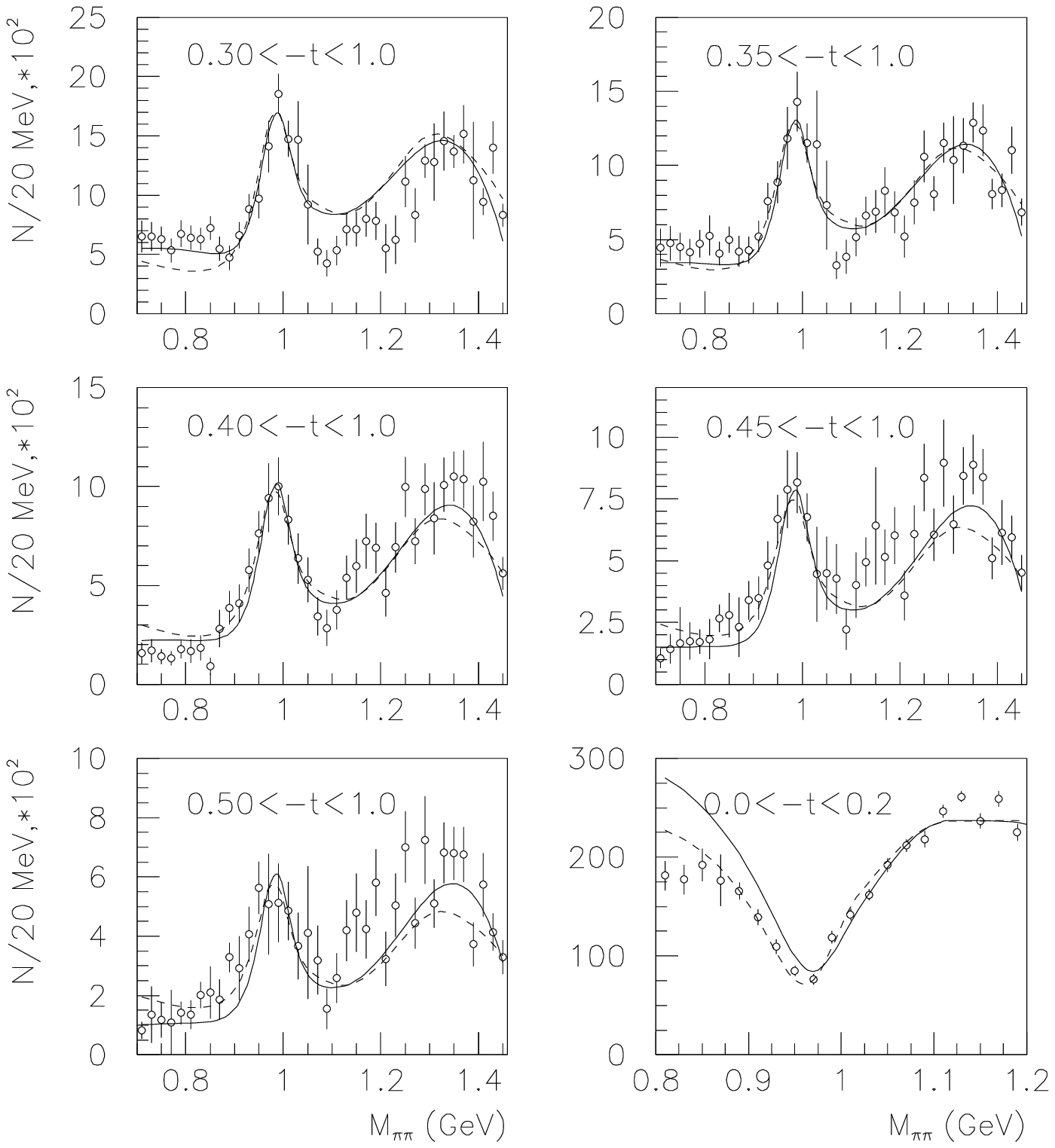,width=14.5cm}\\
Fig. 3. Description of the GAMS data in different $t$-intervals for
coupling parametrization in the form A.
Dashed line shows the  solution
published previously [10] for the fit of GAMS data alone.
\end{center}
\end{figure}

\begin{figure}[hp]
\begin{center}
\epsfig{file=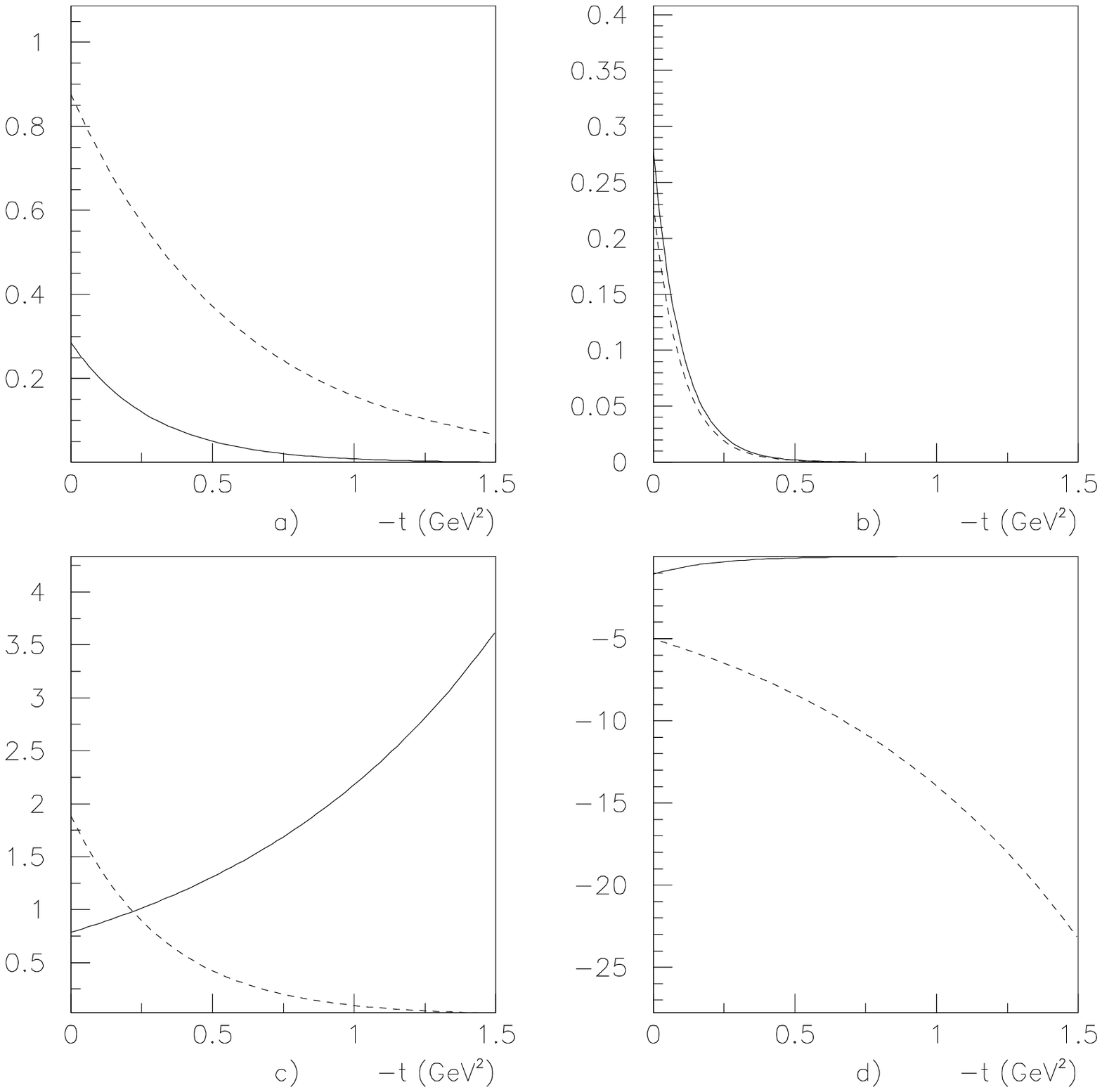,width=14.5cm}\\
Fig. 4 The $t$-dependence of the K-matrix couplings for the
$\pi_{(leading)}$ exchange in the solution A:  a) full curve for
$f_0^{bare}(720)$ and dashed one for $f_0^{bare}(1250)$, b) full curve
for $f_0^{bare}(1600)$ and dashed curve for $f_0^{bare}(1230)$. c), d)
$t$-dependence of the same vertices for the
$a_{1(leading)}$-trajectory exchange (notations are the same as in
Fig. 4a, b).
\end{center}
\end{figure}

\begin{figure}[hp]
\begin{center}
\epsfig{file=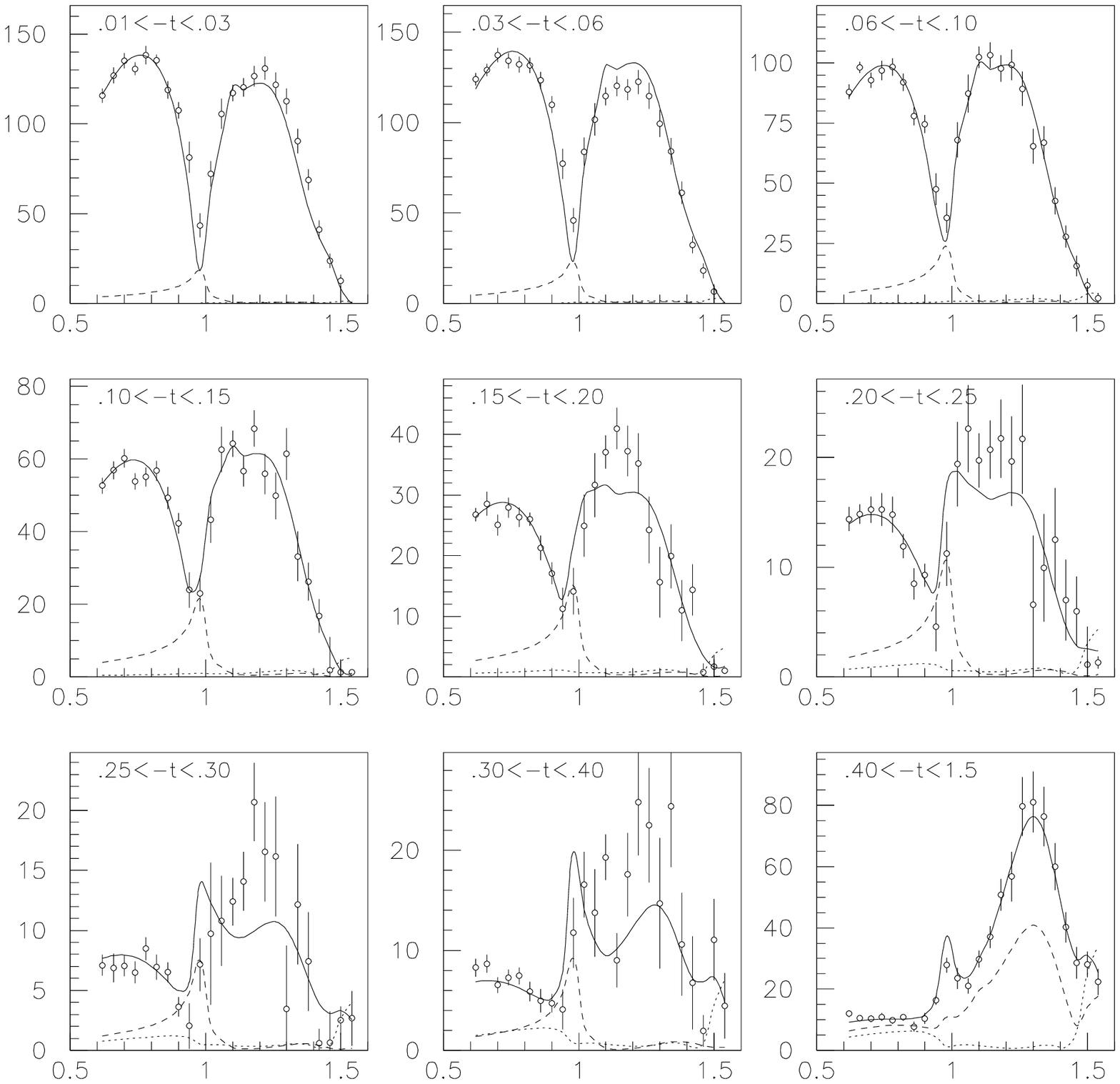,width=14.5cm}\\
Fig. 5. The same as in Fig. 2 but for the solution B1.
\end{center}
\end{figure}

\begin{figure}[hp]
\begin{center}
\epsfig{file=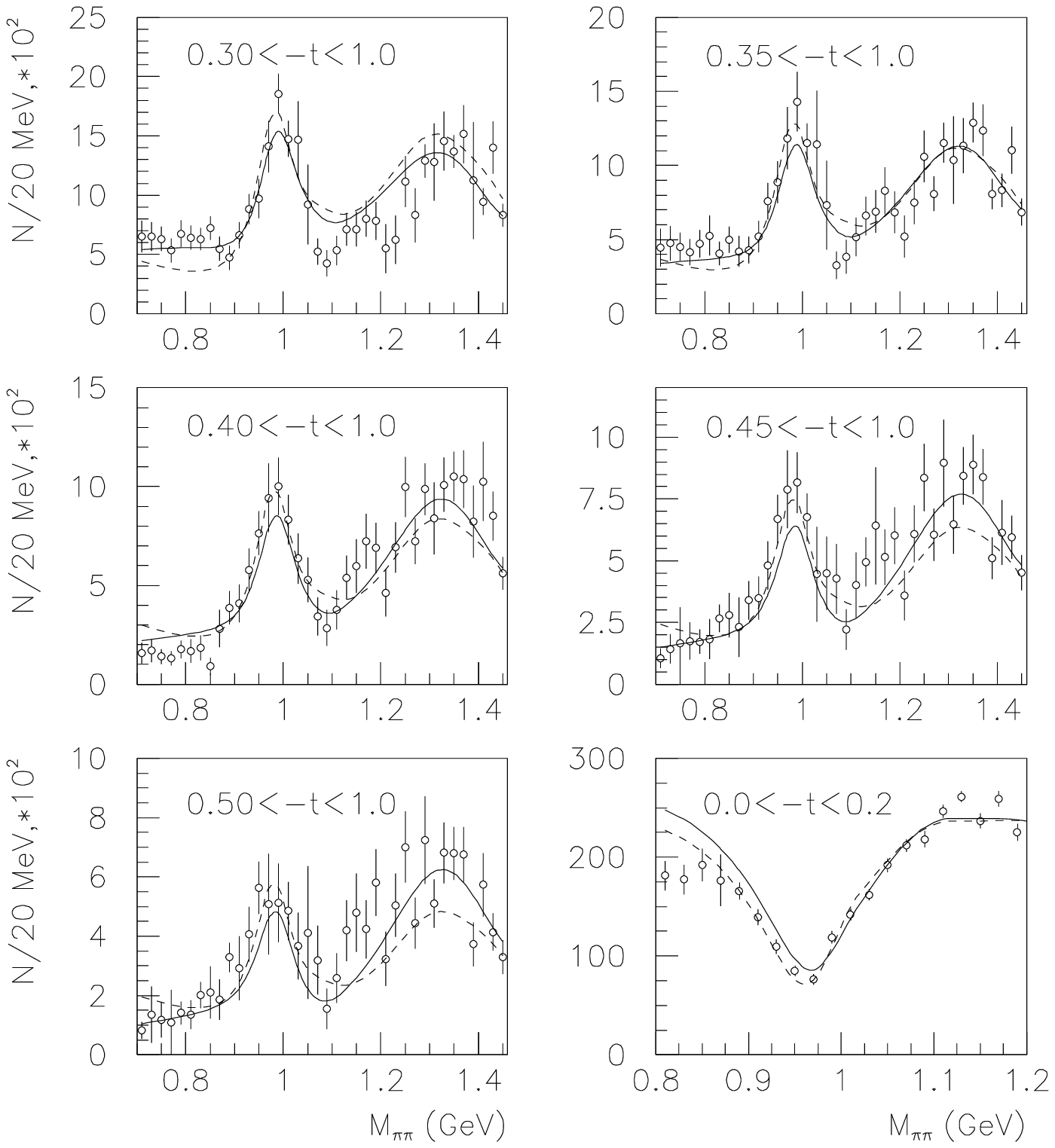,width=14.5cm}\\
Fig. 6. The same as in Fig. 3 but for the solution B1.
\end{center}
\end{figure}

\begin{figure}[hp]
\begin{center}
\epsfig{file=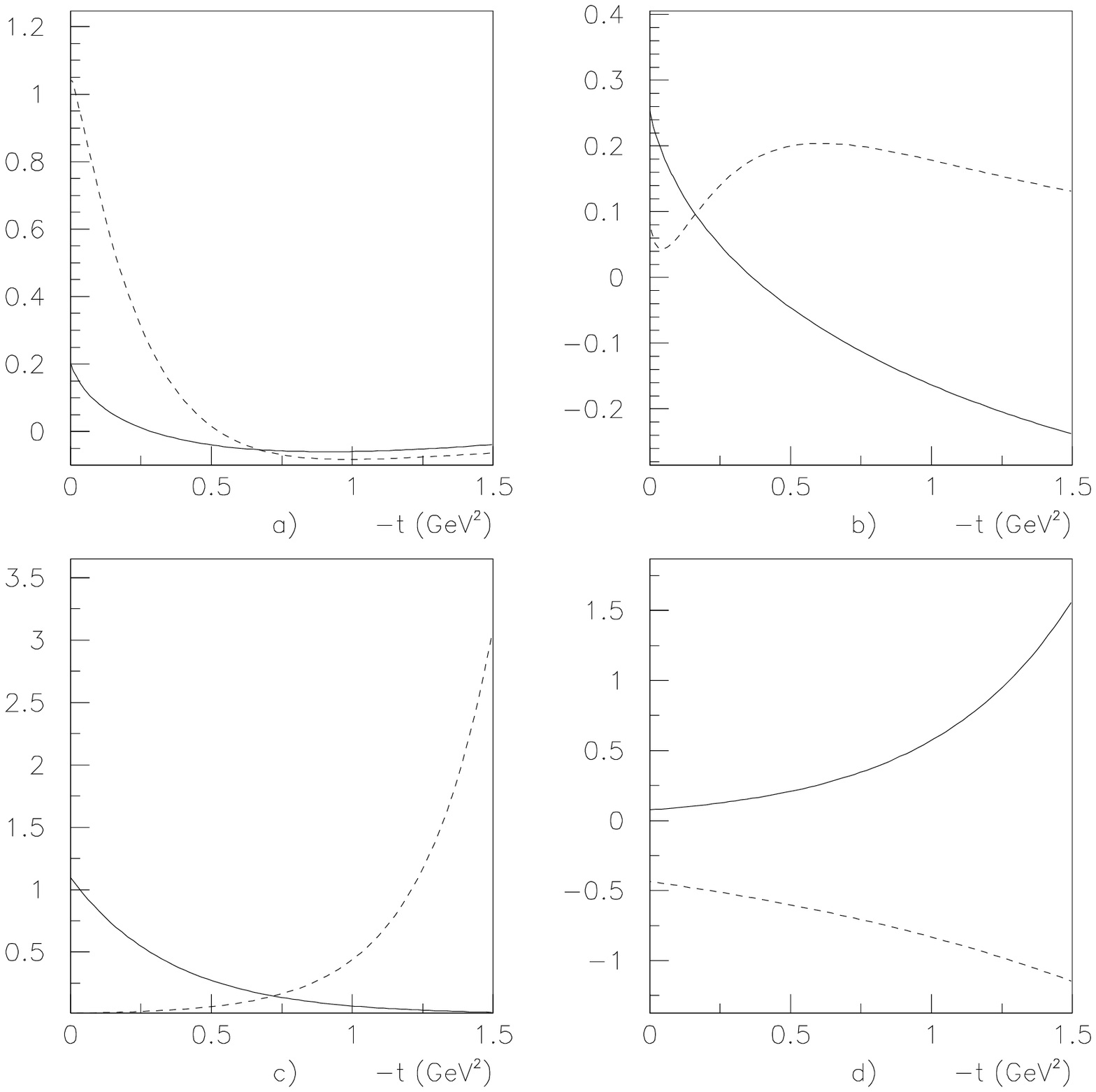,width=14.5cm}\\
Fig. 7. The same as in Fig. 4 but for the solution B1.
\end{center}
\end{figure}

\begin{figure}[hp]
\begin{center}
\epsfig{file=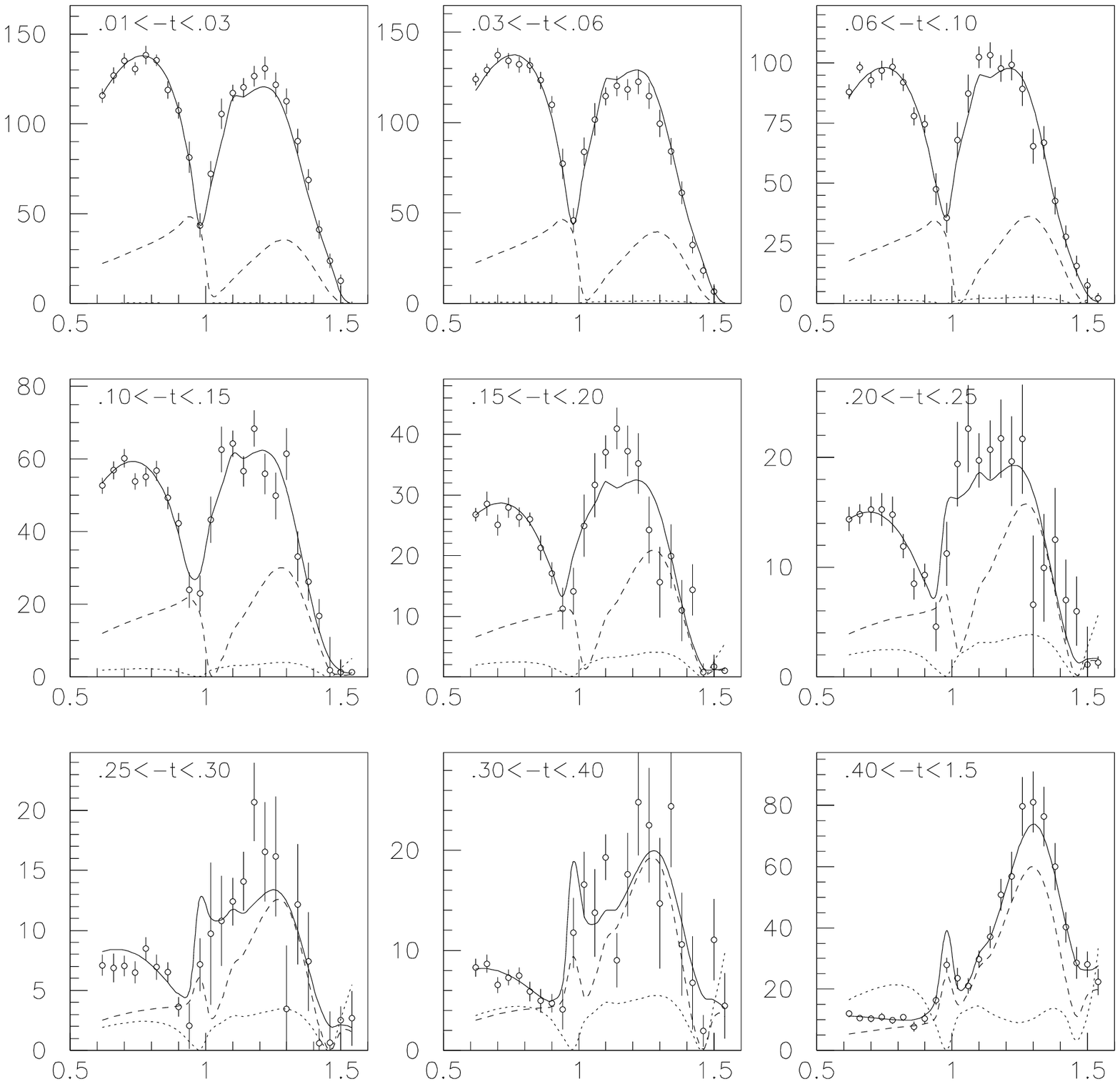,width=14.5cm}\\
Fig. 8. The same as in Fig. 2 but for the solution B2.
\end{center}
\end{figure}

\begin{figure}[hp]
\begin{center}
\epsfig{file=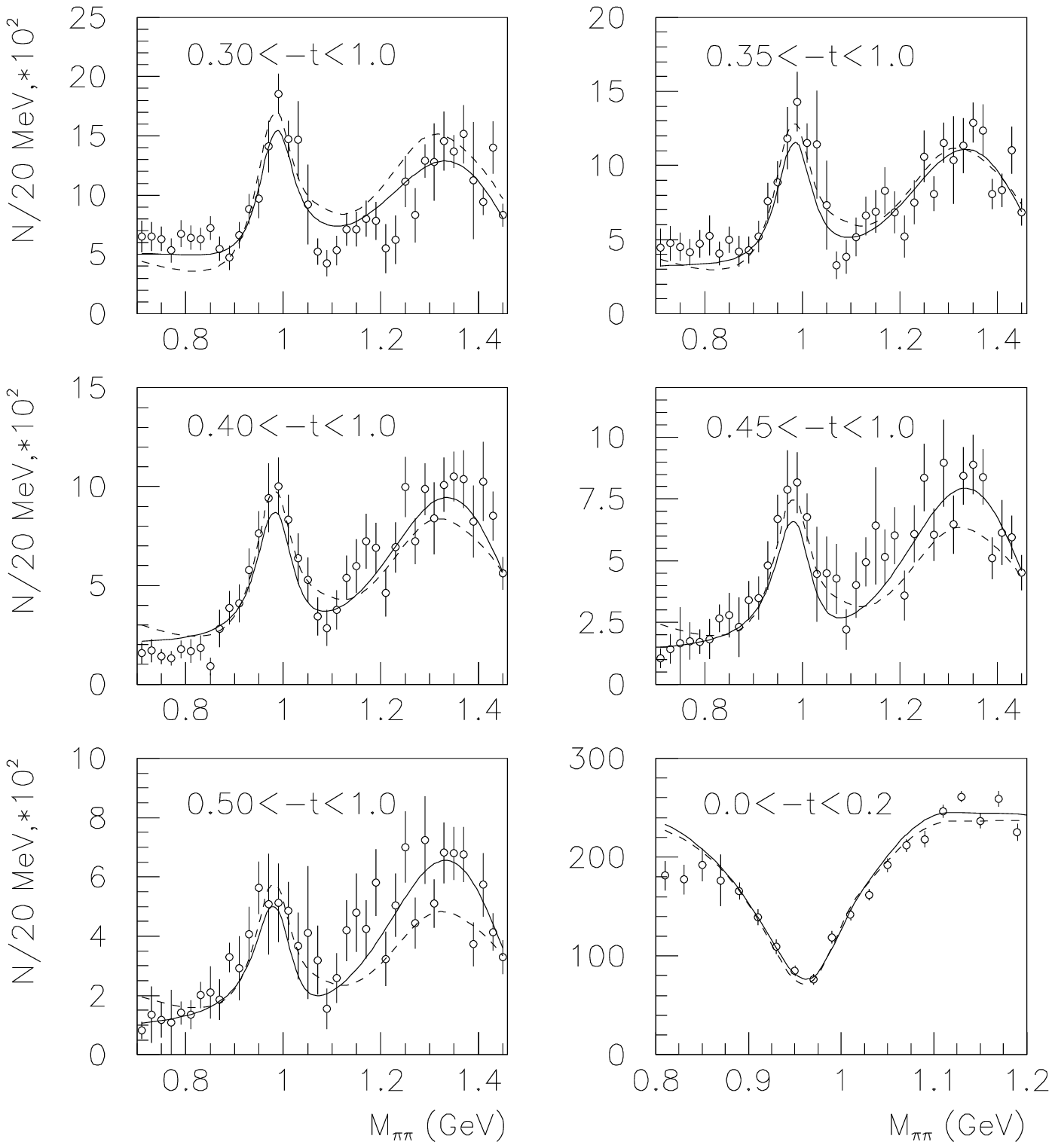,width=14.5cm}\\
Fig. 9. The same as in Fig. 3 but for the solution B2.
\end{center}
\end{figure}

\begin{figure}[hp]
\begin{center}
\epsfig{file=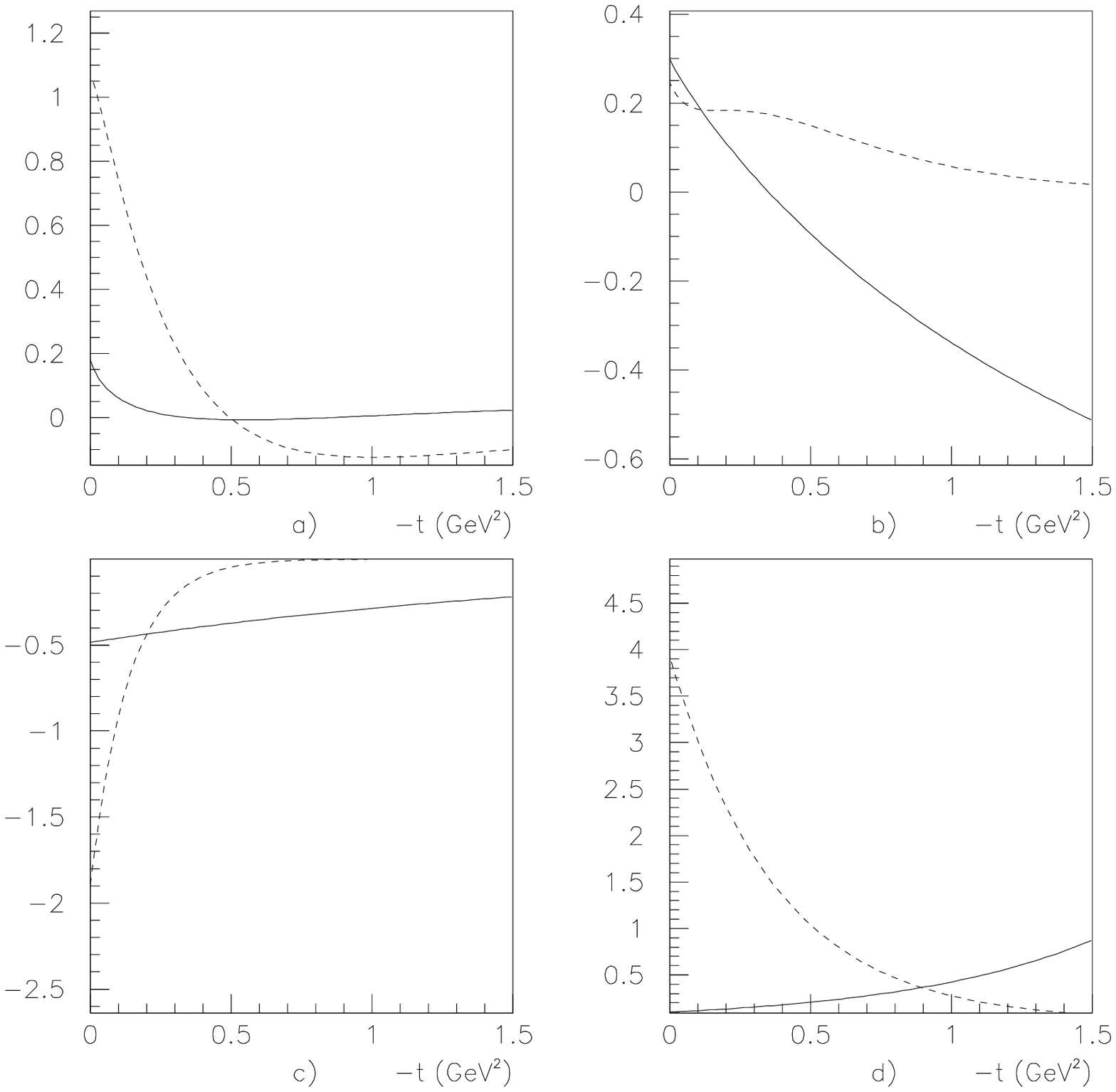,width=14.5cm}\\
Fig. 10. The same as in Fig. 4 but for the  solution B2.
\end{center}
\end{figure}

\end{document}